\documentclass[12pt]{article}
\usepackage{graphicx}
\usepackage{epsfig}
\usepackage[figuresright]{rotating}
\usepackage{amsmath}
\usepackage{amssymb}
\usepackage{amsfonts}
\usepackage{array}

\newcommand{\pibf}{\mbox{\boldmath $\pi$}}
\newcommand{\taubf}{\mbox{\boldmath $\tau$}}

\newcommand{\fpartial}{\mbox{$/\!\!\!\partial$}}

\usepackage{cite}
\usepackage{amssymb}
\usepackage{amsfonts}

\begin{document}

\begin{center}
{\LARGE{\bf Mass generation via the Higgs boson  and the quark condensate of the
QCD vacuum
}}\\[1ex]
 Martin Schumacher\footnote{mschuma3@gwdg.de}\\
II.  Physikalisches Institut der Universit\"at G\"ottingen,
Friedrich-Hund-Platz 1\\ D-37077 G\"ottingen, Germany
\end{center}
\begin{abstract}
The Higgs boson, recently discovered with  a mass of 125.09$\pm$0.24 GeV 
is known to
mediate the masses of elementary particles, but only 2\% of the mass of the
nucleon. Extending a previous investigation  \cite{schumacher14}
and including the strange-quark sector, hadron  masses are derived 
from the  quark condensate of the QCD vacuum and from the  effects of the
Higgs boson. These calculations include the $\pi$ meson, the nucleon 
and the scalar mesons $\sigma(600)$, $\kappa(800)$, $a_0(980)$, $f_0(980)$
and $f_0(1370)$.
The predicted second 
$\sigma$ meson, ${\sigma}'(1344)=|s\bar{s}\rangle$, is investigated and
identified with the $f_0(1370)$ meson. An outlook is given on the hyperons
$\Lambda$, $\Sigma^{0,\pm}$ and $\Xi^{0,-}$.
\end{abstract}

\section{Introduction}

In the standard model the masses of elementary particles  arise from the Higgs 
field acting on  the originally massless particles. 
When applied to the visible matter of the universe this explanation
remains unsatisfactory as long as we consider the vacuum as an empty space.
The QCD vacuum contains a condensate of up and down quarks. Condensate means
that the $q\bar{q}$ pairs are correlated via inter-quark forces
mediated  by gluon exchanges.  As part of the vacuum structure the  $q\bar{q}$
pairs have to be in a scalar-isosclar configuration. This suggests that the
vacuum condensate may be described in terms of a scalar-isoscalar particle,
$|\sigma\rangle =(|u\bar{u}\rangle + |d\bar{d}\rangle)/\sqrt{2}$, providing the
$\sigma$ field. These two descriptions in terms of a vacuum condensate
or  a $\sigma$ field   are essentially equivalent and are the
bases of the Nambu--Jona-Lasinio (NJL) model 
\cite{NJL,lurie64,eguchi76,vogl91,klevansky92,hatsuda94,bijnens96} and the
linear $ \sigma$ model
(L$\sigma$M), \cite{LsigmaM} respectively. Furthermore, it is possible write
down a bosonized 
version of the NJL model where the vacuum condensate is replaced by the
vacuum expectation value of the $\sigma$ field.

In the QCD vacuum the largest part of the mass  $M$ of an originally
massless quark, up (u) or  
down (d), is generated  
independent of the presence of the Higgs field and amounts to $M=326$ MeV
\cite{schumacher14}.
 The Higgs field only adds a small additional part to the 
total constituent-quark mass leading to $m_u= 331$ MeV and
$m_d=335$ MeV for the up and down quark, respectively \cite{schumacher14}.
These constituent quarks 
are the building blocks of the nucleon in a similar
way as the nucleons are in case of nuclei. Quantitatively, we obtain the 
experimental masses of the nucleons after including a binding energy
of 19.6 MeV and 20.5 MeV per constituent quark for the proton and neutron,
respectively, again in analogy to the
nuclear case where the binding energies are 2.83 MeV per nucleon 
for $^3_1$H and 
2.57 MeV per nucleon for $^3_2$He.

In the present work we  extend our previous \cite{schumacher14}
investigation
by exploring in more detail the rules according  to which the  effects
of electroweak (EW) and strong-interaction symmetry breaking combine in order
to generate the masses of hadrons. As a test of the concept, the mass of the
$\pi$ meson is precisely 
predicted on an absolute scale. In the strange-quark sector  the Higgs 
boson is responsible 
for about 1/3 of the constituent-quark mass, so that effects of the interplay
of the two components of mass generation become  essential. Progress is made
by taking into account the predicted second $\sigma$ meson,
$\sigma'(1344)=|s{\bar s}\rangle$ \cite{hatsuda94}. It is found that the
coupling constant of the  $s$-quark coupling 
to the   $\sigma'$ meson  is larger than the corresponding quantity  of the
$u$ and $d$ quarks coupling to 
the $\sigma$ meson by a factor of $\sqrt{2}$. This leads to a considerable
increase of 
the constituent-quark masses  in  the strange-quark sector in comparison with
the 
ones in the non-strange sector already in the chiral limit, i.e.
without the effects  of the Higgs boson. There is  an additional sizable 
increase of the mass generation mediated by the Higgs boson due to a $\sim$ 20 
times stronger
coupling of the $s$ quark to the Higgs boson 
in comparison to the $u$ and $d$ quarks.

In addition to the progress made in \cite{schumacher14} as described above 
that paper contains a History of the subject from 
Schwinger's seminal work of 1957 \cite{schwinger57} to the discovery of the
Brout-Englert-Higgs 
(BEH) mechnism, with emphasis on the Nobel prize awarded to  Nambu in 2008. 
This is the reason  why paper \cite{schumacher14}
has been published as a supplement of  the Nobel lectures of Englert
\cite{englert14} and Higgs \cite{higgs14}.

\section{Symmetry breaking in the non-strange  sector} 
 
\begin{figure}[h]
\begin{center}
\includegraphics[width=0.6\linewidth]{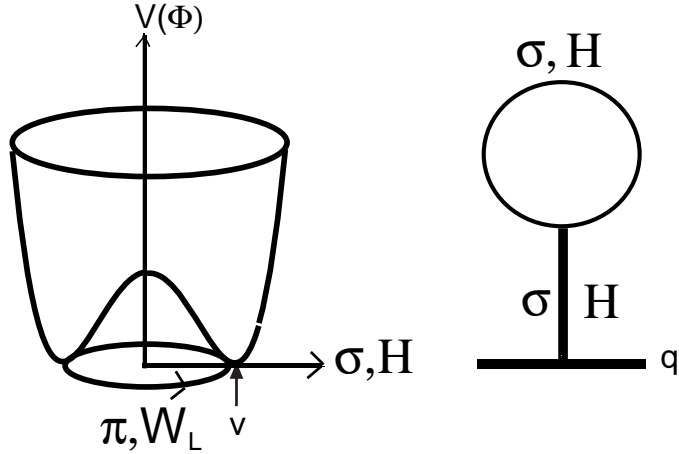}
\end{center}
\caption{Strong-interaction and EW interaction 
symmetry breaking. Left
panel: The L$\sigma$M together with the EW  counterpart. Right panel:
The bosonized NJL model together with the EW  counterpart. }
\label{sombr6}
\end{figure}

In Figure \ref{sombr6} the symmetry breaking process is illustrated. 
The left panel corresponds to the L$\sigma$M, the right panel to the bosonized
NJL
model, together with their EW counterparts. In the left panel 
symmetry breaking provides us with a vacuum
expectation $v_H$ of the Higgs field H and $v_\sigma$ of the $\sigma$-field. 
Without the effects of the Higgs field the strong-interaction  Nambu-Goldstone 
bosons, $\pi$, are massless. The $\pi$ mesons  generate mass via  
the interaction
with the Higgs field  in the presence of the QCD quark condensate, 
as will be outlined
below. The EW counterparts of the $\pi$ mesons are the longitudinal 
components $W_l$ of the weak
vector bosons $W$. These  longitudinal components are transferred into the 
originally  massless weak
vector bosons W via the Brout-Englert-Higgs (BEH) mechanism.
In the right panel the view of the bosonized 
NJL model is presented. The originally
massless quarks interact via the exchanges of a $\sigma$ meson
or a Higgs boson  with the respective $\sigma$ field of the QCD  vacuum 
or the Higgs field  of the EW vacuum. As long
as we consider the two symmetry breaking processes separately we can
write down \cite{schumacher14}
\begin{eqnarray}
&& M=g \,v^{\rm cl}_\sigma,\label{sym1}\\
&& m^0_u=2.03\times 10^{-5}\, v_H, \label{sym2}\\
&& m^0_d=3.66 \times 10^{-5}\, v_H. \label{sym3}
\end{eqnarray}
The quantity g is the quark-$\sigma$ coupling constant which has been derived
to be $g=\frac{2\pi}{\sqrt{3}}$. This quantity leads via Eq. (\ref{sym1})
to the  constituent-quark 
mass
 $M=326$ MeV in the chiral limit (cl),
i.e. without the effects of the Higgs boson. The quantity  $v^{\rm cl}_\sigma
\equiv 
f^{\rm cl}_\pi=89.8$ 
MeV is  the pion decay constant in the chiral limit, serving as vacuum
expectation 
value $(v_\sigma)$. The quantity  $v_H=246$ GeV is the vacuum
expectation value of the Higgs field, leading to the current-quark masses
$m^0_u=5$ MeV and $m^0_d=9$ MeV. These values are the well known current-quark
masses entering into low-energy QCD via explicit symmetry breaking. The
coupling constants in Eqs. (\ref{sym2}) and (\ref{sym3}) are chosen such
that these values are reproduced.

In the following we study the laws according to which the two sources of 
mass generation combine in order to generate the observable particle
masses. For this purpose we write down the well known NJL equation and refer
to \cite{schumacher14} for more details
\begin{equation}
{\cal L}_{\rm NJL }={\bar \psi}(i\fpartial
-m_0)\psi+\frac{G}{2}[({\bar\psi}\psi)^2 
+({\bar \psi} i \gamma_5 \taubf \psi)^2]. \label{NJL} 
\end{equation}
In Eq. (\ref{NJL}) the interaction between fermions is parameterized by the
four-fermion interaction constant G. Explicit symmetry breaking  as mediated
by the Higgs boson is represented by the trace of the 
current-quark mass matrix $m_0$  \cite{vogl91}. From Eq. (\ref{NJL}) the
constituent-quark 
mass in the chiral limit (cl) may be derived via the relation
\begin{equation}
M=G |\langle {\bar \psi}{\psi} \rangle|^{\rm cl}. \label{MG}
\end{equation}
The bosonization of Eq. (\ref{NJL})
is obtained by replacing G by  a propagator
\begin{equation}
G \to \frac{g^2}{(m^{\rm cl}_\sigma)^2- q^2 }, \quad q^2 \to 0
\label{Gtog}
\end{equation}
where $m^{\rm cl}_\sigma$ is the mass of the $\sigma$ meson in the chiral
limit and
   $q$  the momentum carried by the $\sigma$ meson,
and by introducing the $\sigma$ and $\pi$ fields via
\begin{equation}
\sigma=-\frac{G}{g}{\bar \psi}\psi, \quad {\pibf} =-\frac{G}{g}{\bar
  \psi}\,i\,\gamma_5\,{\taubf} \psi. \label{fields}
\end{equation}
Using the Nambu relation $m^{cl}_\sigma=2\,M,$ the quark-$\sigma$ coupling
constant $g=\frac{2\pi}{\sqrt{3}}$,  $M=g\,f^{\rm cl}_\pi$ and 
$f^{\rm cl}_\pi=89.8 \,\,{\rm MeV}$
we arrive at 
\begin{equation}
|\langle {\bar \psi}\psi\rangle|^{\rm cl} =\frac{8\pi}{\sqrt{3}}(f^{\rm
  cl}_\pi)^3 
=(219\,\, {\rm MeV})^3, \quad  G=\frac{1}{4(f^{\rm cl}_\pi)^2}=3.10\times
10^{-5}\,{\rm MeV}^{-2}.
\label{condensatelimit}
\end{equation}
In the present case of small current-quark masses 
it is straightforward to arrive at a version which
includes the effects of the Higgs boson by replacing $f^{\rm cl}_\pi$
with $f_\pi$. The result is  given by
\begin{equation}
|\langle {\bar \psi}\psi\rangle|=\frac{8\pi}{\sqrt{3}}(f_\pi)^3
=(225\,\, {\rm MeV})^3, \quad f_\pi= (92.43\pm 0.26)\,{\rm MeV}. 
\label{condensate}
\end{equation} 
This equation shows that a representation of the QCD quark condensate
through the vacuum expectation value of the $\sigma$ field is possible 
and leads to a  prediction for a value of the vacuum condensate which is
tested and found valid in the next subsection.

Fig. 1 suggests a formal similarity  of strong-interaction and EW
interaction symmetry breaking. This formal similarity is incomplete
because at the present state of our knowledge
we have to consider the Higgs boson as elementary, i.e. without a
fermion-antifermion substructure. Formerly, 
substructures  of the Higgs field
in terms of techniquark-antitechni\-quark pairs  or top-antitop quark  pairs
have been discussed. In the techniquark-model strong self-coupling has been
discussed leading to a predicted Higgs-boson mass of $\sim 1$ TeV.  In the
top-quark model the predicted Higgs boson mass is expected to be $\sim 2m_t$.
With a comparatively small experimental mass of $m_H=125.1$ GeV of the Higgs 
boson these models seem to be
excluded. In case of the left
panel we may write down a potential in the form \cite{halzen84}
\begin{equation}
V(\phi)=-\mu^2\phi^\dagger\phi+\lambda(\phi^\dagger\phi)^2,\quad \mu^2>0,\quad
\lambda >0.
\label{potential}
\end{equation}
The potential $V(\phi)$ then has its minimum at a finite value of $|\phi|$
where
\begin{equation}
\phi^\dagger\phi=\frac12 (\phi^2_1+\phi^2_2+\phi^2_3+\phi^2_4)=
\frac{\mu^2}{2\lambda}.
\label{minimum-1}
\end{equation}
We can choose, say,
\begin{equation}
\phi_1=\phi_2=\phi_4=0,\quad \langle\phi_3\rangle=
\sqrt{\frac{\mu^2}{\lambda}}\equiv v_H
\label{minimum-2}
\end{equation}
where $v_H=246$ GeV is the vacuum expectation value of the Higgs field. In the
case of strong interaction the corresponding relations are
\begin{equation}
\phi^\dagger\phi=\frac12 (\sigma^2 + \pibf^2)=\frac{\mu^2}{2\lambda}
\label{minimum-3}
\end{equation}
with
\begin{equation}
\pibf=0, \quad \langle \sigma\rangle=\sqrt{\frac{\mu^2}{\lambda}}\equiv
v^{\rm cl}_\sigma\equiv f^{\rm cl}_\pi. \label{minimum-4}
\end{equation}

It is of interest to compare the selfcoupling strengths of strong interaction
symmetry breaking with the one of EW  symmetry breaking. The sigma-meson
mass in the chiral limit   may be expressed in two ways \cite{schumacher14}
\begin{equation}
m^{cl}_\sigma=2\,g\,v^{\rm cl} _\sigma, \quad m^{\rm
  cl}_\sigma=\sqrt{2\,\lambda_\sigma} 
\,v^{\rm cl}_\sigma\label{minimum-5}
\end{equation}
where the first version corresponds to the NJL model and the second to the
L$\sigma$M. With $v_\sigma=f^{\rm cl}_\sigma=89.8\,\,{\rm MeV}$ and
$g=\frac{2 \pi}{\sqrt{3}}=3.63$ this leads to
\begin{equation}
m^{\rm cl}_\sigma=652 \,\,{\rm MeV}\,\, {\rm and}\,\, 
\lambda_\sigma=\frac{8\pi^2}{3}
=26.3. \label{minimum-6}
\end{equation}
For the Higgs boson we have 
\begin{equation}
m_H=\sqrt{2\lambda_H}\,v_H,\,\,m_H=125.1\,\,{\rm GeV}\,\,{\rm and\,\,} v_H=246
\,\,{\rm GeV}\label{minimum-7}
\end{equation} 
leading to
\begin{equation}
\lambda_H=0.130. \label{minimum-8}
\end{equation}
We see that the  strong-interaction selfcoupling is a factor 
$\lambda_\sigma/\lambda_H=202$
larger than the EW selfcoupling. But except for this, there indeed
is a formal similarity between the two versions of symmetry breaking. This
may help  to get a better understanding of the underlying physics of the 
Higgs boson.

\subsection{Prediction of the masses of the $\pi$ and the $\sigma$ 
meson}

The mass of the $\pi$ meson is given by the Gell-Mann--Oakes--Renner (GOR)
relation in the form
\begin{equation}
m^2_\pi f^2_\pi=(m^0_u + m^0_d)|\langle {\bar\psi}\psi \rangle|,\quad
|\langle \bar{\psi}\psi\rangle|=\frac12|\langle \bar{u}u + \bar{d}d\rangle|
=\frac{8\pi}{\sqrt{3}}(f_\pi)^3,
\label{GOR1}
\end{equation} 
(see Eq. (\ref{condensate}) leading to
\begin{equation}
m^2_\pi =(m^0_u + m^0_d)\frac{ 8 \pi}{\sqrt{3}}f_\pi.
\label{GOR2}
\end{equation}
The numerical value derived from Eq. (\ref{GOR2}) is
$m_\pi=137.0$ MeV, to be compared with the experimental values $m^0_\pi=
135.0$ MeV and  
$m^{\pm}_\pi= 139.6$ MeV. Apparently, the predicted value of  the 
$\pi$ meson mass is quite satisfactory when use is made of the current-quark
masses following from Eqs. (\ref{sym2}) and (\ref{sym3}). Furthermore, in the
neutral 
$\pi^0$ we find a Coulomb 
attraction between the quark and the antiquark leading to an
experimental value smaller than the predicted value, whereas in the charged
$\pi^{\pm}$ we find a Coulomb repulsion between the quark and the antiquark
leading to an experimental value larger than the predicted value.

From Eqs. (\ref{sym1}) -- (\ref{sym3}) we can see that the constituent quark
in the chiral limit and the current quarks generate masses independently. 
But these masses 
are not the ones observed in low-energy QCD. The procedure to arrive at
predictions for
the observable masses has been derived from arguments given by
the L$\sigma$M as well as NJL models. First we calculate the mass of the 
$\sigma$ meson in the chiral limit via the Nambu relation
\begin{equation}
m^{\rm cl}_\sigma=2M=652 \,\,{\rm MeV} \quad {\rm with}\quad 
M=\frac{2\pi}{\sqrt{3}}f^{\rm cl}_\pi=326\,\,{\rm
    MeV} \label{NambuEq} 
\end{equation}
and then use the relation
\begin{equation}
m_\sigma=\left(4M^2+{{\hat m}_\pi}^2\right)^{1/2}=666 \,\,{\rm
  MeV}.  \label{sigma-mass} 
\end{equation}
In Eq. (\ref{sigma-mass}) the effects of the Higgs boson enter via the average
pion mass ${\hat m}_\pi$. It is interesting to note that in the limit of
small current-quark masses the result of Eq. (\ref{sigma-mass}) can also be
derived by simply adding the contributions from the QCD quark-condensate
and from the Higgs boson, leading to
\begin{eqnarray}
&&m_u=M+m^0_u=331\,\,{\rm MeV}, \label{up}\\
&&m_d=M+m^0_d= 335\,\,{\rm MeV,} \label{down}\\
&&m_\sigma=m^{\rm cl}_\sigma+ m^0_u + m^0_d= 666\,\,{\rm MeV}. \label{sigmass}
\end{eqnarray}
The arguments leading to an equivalence of Eqs. (\ref{sigma-mass}) and 
(\ref{sigmass}) are as follows. Eq. (\ref{sigma-mass}) can be written
is the form 
\begin{equation}
m_\sigma=m^{\rm cl}_\sigma+\frac12\frac{\hat{m}^2_\pi}{m^{\rm cl}_\sigma}
+\cdots 
=m^{\rm cl}_\sigma + (m^0_u + m^0_d).
\label{sigmass1}
\end{equation}
Here use has been made of
Eqs. (\ref{GOR2}) and (\ref{NambuEq}) to show that the term 
$\frac12\frac{\hat{m}^2_\pi}{m^{\rm cl}_\sigma}$
of Eq. (\ref{sigmass1}) is  equal to $(m^0_u+m^0_d)$
if higher order terms amounting to 4\% and the deviation of 
$f^{\rm cl}_\pi/f_\pi$ from 1 amounting to 3\% are neglected.

The result shown in Eq. (\ref{sigmass}) is  in line with the  expectation 
that the $\sigma$ meson
is a loosely bound object where no additional term, {\it viz.}
the binding energy $B$,
has to be taken in account. 
This is different  in the case of baryons as we will
see later.

\section{The fundamental structure constants magnetic moment, 
polarizability
  and mass of the nucleon} 
In the previous paper \cite{schumacher14} it has been shown that 
 it is possible to make precise predictions 
 on an absolute scale
not only for  the 
mass of the nucleon but also for
such sophisticated structure constants as there are the magnetic moment and the
polarizability. This is an important finding because it implies that the
underlying models are confirmed in different and complementary ways.
In the present article  we return to this problem and make some necessary
amendments. 

\subsection{The magnetic moment of the nucleon}

The magnetic moments of the nucleon are given
by 
\begin{eqnarray}
&&\mu_p=\frac43\mu_u - \frac13 \mu_d, \label{magneticmoment-1}\\
&&\mu_n=\frac43\mu_d - \frac13 \mu_u \label{magneticmoment-2}
\end{eqnarray}
in units of the nuclear magneton $\mu_N=e\hbar/2m_p$. Constituent-quark masses
enter through the relations   
\begin{equation}
\mu_u=\frac23\frac{m_p}{m_u},\quad
\mu_d=-\frac13\frac{m_p}{m_d},\label{magneticmoment-3} 
\end{equation}
where $m_u=331$ MeV and $m_d=335$ MeV. This leads to the magnetic
moments of the constituent quarks
\begin{equation}
\mu_u=1.890, \quad \mu_d= -0.934 \label{magneticmoment-4}
\end{equation}
and to predicted magnetic moments of the nucleon
\begin{equation}
\mu^{\rm theor}_p= 2.831 \quad  \mu^{\rm theor}_n=
-1.875. \label{magneticmoment-5} 
\end{equation}
Comparing these values with the experimental
 magnetic moments of the nucleon
\begin{equation}
\mu^{\rm exp}_p=2.79285,\quad \mu^{\rm exp}_n=-1.91304
\label{magneticmoment--6}
\end{equation}
we arrive at  very small differences $\Delta \mu=\mu^{\rm exp}-\mu^{\rm theor}$
\begin{equation}
\Delta\mu_p=-0.038,\quad \Delta\mu_n=-0.038.
\label{magneticmoment-7}
\end{equation}
Apparently, the necessary corrections to the quark-model predictions
of the magnetic moments for the proton and neutron are the same. This 
may help to find an explanation for these corrections. The most probable 
explanation  may be found in terms of meson exchange currents
though available calculations lead to too large values.

In the present work
we are mainly interested in the  small sizes of $1.4\% - 2.0\%$ of the
differences 
showing that the predictions obtained on the basis of the NJL model
are very precise.

\subsection{The polarizabilities of the nucleon}

A nucleon in an electric field {\bf E} and magnetic field {\bf H}
obtains an electric dipole moment {\bf d} and magnetic dipole moment
{\bf m} given by
\begin{eqnarray}
&& {\bf d}=4\,\pi\,\alpha\,{\bf E} \label{electricpol}\\
&& {\bf m}=4\,\pi\,\beta\, {\bf H} \label{magneticpol}
\end{eqnarray} 
in a unit system where the electric charge $e$ is given by $e^2/4\pi
=\alpha_{\rm em}=1/137.04$. The quantities $\alpha$ and $\beta$ are the
electric and magnetic polarizabilities belonging to the fundamental
structure constants of the nucleon. It is of importance
that these quantities are composed of two components
\begin{eqnarray}
&&\alpha=\alpha^s+\alpha^t, \label{st1}\\
&&\beta=\beta^s + \beta^t \label{st2}
\end{eqnarray}
where the superscript $s$ denotes the $s$-channel contribution and the
superscript $t$ the $t$-channel contribution. The $s$-channel contribution
is related to the meson-photoproduction amplitudes  of the nucleon via the
optical theorem whereas the $t$-channel contribution is related to the
$\sigma$ meson as part of the constituent-quark structure. Therefore,
it is possible to use the polarizabilities as a tool to test the predicted
mass and structure of the $\sigma$ meson. This is summarized in the following
equations \cite{schumacher13}
\begin{eqnarray}
&&|\sigma\rangle=\frac{1}{\sqrt{2}}(|u\bar{u}\rangle+|d\bar{d}\rangle),\quad
{\cal M}(\sigma\to\gamma\gamma)=\frac{\alpha_{\rm em}N_c}{\pi\,f_\pi}
\left[\left(\frac23\right)^2+\left(\frac{-1}{3}\right)^2\right],
\label{sigpol1}    \\
&&(\alpha-\beta)^t_{p,n}=\frac{g_{\sigma NN}{\cal M}(\sigma\to\gamma\gamma)}
{2\pi\,m^2_\sigma}=15.2\,\, ,\quad (\alpha+\beta)^t_{p,n}=0
\label{sigpol2}\\
&&\alpha^t_{p,n}=+7.6\quad\quad\quad\quad\quad\quad\quad
\beta^t_{p,n}-7.6\label{sigpol3}\\
&&\alpha^s_p=+4.5,\quad \alpha^s_n=+5.1,\quad \beta^s_p=+9.4,\quad
\beta^s_n=10.1 \label{sigpol4}
\end{eqnarray}
in units of $10^{-4}{\rm fm}^3$,
where use is made of $g_{\pi NN}=g_{\sigma NN}=13.169\pm0.057$ and 
$m_\sigma=666$ MeV as predicted by the NJL model.

The polarizability components listed in Eq. (\ref{sigpol3}) correspond
to the $t$-channel and have been calculated from Eqs. (\ref{sigpol1}) and
(\ref{sigpol2}). The polarizability components in Eq. (\ref{sigpol4})
correspond to the $s$-channel and have been calculated from high-precision 
meson photoproduction amplitudes \cite{schumacher13}. 
\begin{table}[h]
\caption{Total predicted polarizabilities and experimental results (unit 
$10^{-4}$fm$^3$)} 
\begin{tabular}{l|ll|ll}
\hline
&$\alpha_p$&$\beta_p$&$\alpha_n$&$\beta_n$\\
\hline
total predicted&+12.1&+1.8&+12.7&+2.5\\
experim.  result  & +$(12.0\pm 0.6)$&+$(1.9\mp 0.6)$&+$(12.5\pm 1.7)$&+$(2.7\mp
1.8)$\\ 
\hline
\end{tabular}
\label{tab1}
\end{table}

The purpose of this subsection is to show that the $q\bar{q}$ structure of the
$\sigma$ meson as given in Eq. (\ref{sigpol1}) together with the mass
$m_\sigma= 666 $ MeV leads to an excellent agreement 
of the predicted 
with the experimental
polarizabilities of the nucleon as shown in Table \ref{tab1}. Furthermore,
there is a Compton-scattering 
experiment on the proton where the $\sigma$ meson as part of the
constituent-quark structure 
is directly visible in the differential cross section for Compton scattering
in the energy range from 400--700 MeV and at large scattering angles
\cite{galler01,wolf01,schumacher10}. This
latter experiment leads to a $\sigma$ meson mass of $m_\sigma= 600\pm 70$ 
 MeV \cite{schumacher13} in good agreement with the standard value
 $m_\sigma=666$ MeV.
  
\subsection{The masses of the nucleons}

As shown above, the constituent-quark masses including the effects of the
Higgs boson are
$$ m_u= 331\,\, {\rm MeV}\quad {\rm and} \quad
m_d= 335\,\, {\rm MeV}.$$
This leads to the nucleon masses
\begin{eqnarray}
&&m^0_p=2 m_u +m_d= 997 {\rm MeV}, \label{masses 3}\\
&&m^0_n=2 m_d +m_u= 1001  {\rm MeV}. \label{masses 4}
\end{eqnarray}
The difference of these quantities from the experimental values
\begin{eqnarray}
&& m_p= 938.27 \,\,{\rm MeV}, \label{masses5}\\
&& m_n = 939.57 \,\,{\rm MeV}, \label{masses6}
\end{eqnarray}
may be interpreted in terms of a binding energy $B$, leading to 
\begin{eqnarray}
&& B_p= m^0_p -m_p = 59 \,\,{\rm MeV}, \label{masses7}\\
&& B_n = m^0_n - m_n = 61 \,\,{\rm MeV}. \label{masses8}
\end{eqnarray}
The larger binding energy $B_n$ of the neutron compared to $B_p$ of the
proton, $B_n-B_p\approx 2$ MeV, has previously \cite{schumacher14} been
interpreted in terms of a Coulomb 
attraction, which leads  to zero in case of the proton but to a nonzero value
of the right order of magnitude in case of the neutron. The arguments were
as follows. The electromagnetic potential acting between three constituent
quarks may be written in the form
\begin{equation}
U=\sum_{i<j}\frac{e_ie_j}{r_{ij}}\alpha_{em}\hbar c
\label{empotential}
\end{equation}
where the denominator has been  replaced by an educated guess for the
average interquark distance  
\cite{schumacher14}, 
{\it viz.}  $\langle r_{ij} \rangle\approx 0.3$ fm \cite{schumacher14}. This
tentative 
consideration leads to $U_p=0$ MeV and  $U_n= -1.6 $ MeV or $B_n -B_p =1.6$
MeV. The difference 
$B_n - B_p=2.0$ MeV contained in Eqs. (\ref{masses7}) and (\ref{masses8})
would lead to $\langle r_{ij} \rangle =0.24$ fm in reasonable agreement with
the educated guess.

It may be expected that a calculation of the hadronic binding energy 
of the nucleon leads to interesting insights into the constituent-quark
structure of the nucleon. At the present point of research we leave this
as an open problem for further investigations.

\section{Hadron masses in the  SU(3) sector}

In the SU(2) sector we have the $\pi$ mesons  serving as Nambu-Goldstone
boson and the  $\sigma(666)$ meson serving as Higgs boson of strong
interaction. In the SU(3) sector we expect an octet $\pi$, $K$ and  $\eta$
of Nambu-Goldstone bosons and a nonet $\sigma(666)$, $\kappa(800)$
and $f_0(980),a_0(980)$ of Higgs bosons of strong interaction. This latter
case has been investigated in  a previous paper \cite{schumacher11}.  
Since the $\sigma(666)$ meson is given by the
$|n\bar{n}\rangle=|(u\bar{u}+d\bar{d}\rangle/\sqrt{2}$ state one should 
expect that the $f_0(980)$ meson is given by the related $|s\bar{s}\rangle$ 
state. This, however cannot be the case because the mesons $f_0(980)$
and $a_0(980)$ have equal masses and, therefore, must have 
an equal fraction $f_s$ of strange quarks in the meson structure. There are arguments
that the missing $|s\bar{s}\rangle$ scalar meson may be  identified with the
$f_0(1370)$ state. This has previously been pointed out by Hatsuda and
Kunihiro 
\cite{hatsuda94} and recently by Fariborz et al. \cite{fariborz15}. According
to Hatsuda and Kunihiro the $f_0(1370)$ may be considered as a second sigma
meson, $\sigma'$, which takes over the r\^ole of the $\sigma$ meson
when we replace
$|n{\bar n} \rangle$ by $|s{\bar s}\rangle$. The mesons $\sigma$ and $\sigma'$
differ by the fact that the wave function of the $\sigma$ meson contains
two flavors, $N_f=2$, whereas the wave function of $\sigma'$ contains only
one flavor, $N_f=1$. This may lead to the consequence that the coupling constant
of a $u$ or $d$ quark to the $\sigma$ meson may  be different from the
coupling strength of a $s$ quark to the $\sigma'$ meson. This point will
be investigated in the next subsection.

\subsection{Properties of the strange-quark $\sigma$ meson
in the chiral limit}

The nonstrange and the strange-quark $\sigma$ mesons differ by the fact that
we have flavor numbers $N_f=2$ for $\sigma$  and $N_f=1$ for $\sigma'$. Due to
this  difference it may
be expected that these two mesons have different quark-mesons coupling
constants $g$ for the $u$ and $d$ quarks coupling  to the
$\sigma$ meson and the coupling constant $g_S$ for the of the
$s$ quark  to the $\sigma'$ meson. This difference 
has been investigated by Delbourgo and Scadron 
\cite{delbourgo95,delbourgo98} (see also \cite{schumacher06})
on the basis of a diagrammatic approach. 
 Using dimensional regularization the  graphs sum up in the chiral limit to
\cite{delbourgo98}
\begin{eqnarray}
&&(m^{\rm cl}_{\sigma })^2=16\,i\,N_c\,g^2\int\frac{d^4p}{(2\pi)^4}\left[
\frac{M^2}{(p^2-M^2)^2}-\frac{1}{p^2-M^2}\right]
=\frac{N_c\,g^2\,M^2}{\pi^2}, \label{nonstrane}\\
&&(m^{\rm cl}_{\sigma S})^2=8\,i\,N_c\,g^2_{S}\int\frac{d^4p}{(2\pi)^4}\left[
\frac{M^2_{S}}{(p^2-M^2_{S})^2}-\frac{1}{p^2-M^2_{S}}\right]
=\frac{N_c\,g^2_{S}\,M^2_{S}}{2\pi^2}. \label{strane}
\end{eqnarray}
Here use has been made of a $\Gamma$ function identity 
$\Gamma(2-l)+\Gamma(1-l)\to -1$ in $2l=4$ dimensions.
The conclusions drawn from these considerations are
\begin{equation}
g=\frac{2 \pi}{\sqrt{3}},\quad g_S=\sqrt{2}g,\quad m^{\rm cl}_S= 2M_S\,\, {\rm
  and}\,\, M_S=\sqrt{2} M. 
 \label{conclusions}
\end{equation}
The last of the relations in Eq. (\ref{conclusions}) implies that
the vacuum expectation values of the nonstrange and strange-quark
sigma fields are the same in the chiral limit. Keeping this in mind
we arrive at
\begin{equation} 
M_S=461\,\, {\rm MeV}\quad {\rm and}\quad m^{\rm cl}_{\sigma S}=922\,\,{\rm  MeV}.
\label{conclusions2}
\end{equation}

\subsection{Current-quark mass of  the $s$ quark}

In the SU(3) sector the effects of the Higgs boson enter into the mass 
generation process
via the current-quark masses of the $u$, $d$ and $s$ quarks.  In 
low-energy  QCD 
the current-quark masses of the $u$ and $d$ quarks
are well known to be $m^0_u=5$ MeV and $m^0_d=9$ MeV, as already stated above.  
The current-quark mass of the $s$ quark is less well known and, therefore,
requires some further investigation. Here we first make the attempt to exploit
an analog of  Eq. (\ref{GOR2}) given for $\pi$ meson and   write down for the
$K^+$ 
\begin{equation}
m^2_{K^+}=(m^0_u+m^0_s)\frac{8\pi}{\sqrt{3}}f_K. \label{GORK}
\end{equation}
This equation allows to calculate the current-quark mass $m^0_s$
from the mass of the $K^+$ meson, the $K^+$ meson decay constant $f_K$
and the current-quark mass $m^0_u$. 
Then with $m_{K^+}=493,67$ MeV,    $f_K=110.45$ MeV, $m^0_u=5$ MeV we arrive
at $m^0_s=147$ MeV.

For the case of low-energy QCD  the following values
may be found in the literature:
$m^0_s=(161\pm 28)$ MeV \cite{hatsuda94,narison87}, $m^0_s=(175\pm 55)$ MeV
\cite{gasser82}
$m^0_s=(199\pm 33)$ MeV \cite{hatsuda94,dominguez87}.  These data span
the range from $m^0_s=133$ to $m^0_s= 232$ MeV with the value $m^0_s=147$ MeV
following from Eq. (\ref{GORK}) being close to the lower limit.
The following considerations appear to be justified. EW interaction
alone should lead to a definite value, $m^0_s({\rm EW})$, of the current-quark
mass of the strange quark and deviations from this value may be due to an
incomplete decompositions of the effects of EW and strong interaction.
 Eq.  (\ref{GORK}) contains effects of strong-interaction
 only due to the decay constant $f_K$ which is well determined
experimentally. Furthermore, the relation  Eq. (\ref{GORK}) is well
justified through its close similarity with the corresponding 
Eq. (\ref{GOR2}) derived and found valid for the $\pi$ meson. This
consideration leads to the supposition that a  value around $m^0_s=147$ MeV
may  be identified with $m^0_s({\rm EW})$.

\subsection{The masses and structures of scalar mesons}

In the SU(3) sector it has become customary to distinguish between
scalar mesons with masses below 1 GeV and scalar mesons with masses above
1 GeV. The properties of the scalar mesons with masses below 1 GeV have been
investigated and described in a previous paper \cite{schumacher11}. 
There is a nonet of scalar mesons  with a 
$(q\bar{q})^2$ tetraquark structure-component coupled to a $q\bar{q}$
component. The reason for the 
assumption of a tetraquark structure-component is that in a $q\bar{q}$
model the electrically neutral $a_0(980)$ meson should have a quark structure
in the form $(-u\bar{u}+d\bar{d})/\sqrt{2}$ and the $f_0(980)$ meson 
a quark structure in the form $s\bar{s}$. This would lead to the consequence
that the masses should be different, whereas in reality they are
equal to each other. On the other hand in a tetraquark model the fraction
of strange quarks $f_s$ is equal in the two mesons as can be seen 
in  Table
\ref{tab:tetraquark}.
\begin{table}[h]
\caption{Summary of scalar mesons in the $(q\bar{q})^2$ representation 
according to \cite{jaffe78}. $Y$: hypercharge, $I_3$: isospin component,
$f_s$: fraction of strange and/or anti-strange quarks in the tetraquark
structure.}
\setlength{\extrarowheight}{4pt} 
\begin{center}
\begin{tabular}{lcccccll}
\hline
$Y/I_3$ & -1 &-1/2 & 0 & +1/2 & +1& Meson & $f_s$\\
\hline
$+1$&&$d{\bar s}u{\bar u}$&&$u{\bar s}d{\bar d}$&&$\kappa(800)$&1/4\\
0&&&$u{\bar d}d{\bar u}$&&& $\sigma(600)$&0\\
0&$d{\bar u}s{\bar s}$& &$s{\bar s}(u\bar{u}-d {\bar d})/\sqrt{2}$&&
$u{\bar d}s{\bar s}$&$a_0(980)$& 1/2\\
0&&&$s{\bar s}(u\bar{u}+d {\bar d})/\sqrt{2}$&&&$f_0(980)$&1/2\\
-1&&$s{\bar u}d{\bar d}$&&$s{\bar d}u{\bar u}$&&$\kappa(800)$&1/4\\
\hline
\end{tabular}
\end{center}
\label{tab:tetraquark}
\end{table}

The tetraquark structure of the $\sigma(600)$ meson needs a special
consideration. For this purpose we study the two reaction chains given in
Eqs. (\ref{doorway1}) and (\ref{doorway2})
\begin{eqnarray}
&&\gamma\gamma\rightarrow(u{\bar u} + d{\bar d})/\sqrt{2}\rightarrow u{\bar
  d}d{\bar u} 
\rightarrow \pi\pi, \label{doorway1}\\
&&\gamma\gamma\rightarrow(u{\bar u} + d{\bar d})/\sqrt{2}
\rightarrow N{\bar N}.
\label{doorway2}
\end{eqnarray}
Eq. (\ref{doorway1}) describes 
the two-photon production of a pion pair. The two photons first
excite the $q{\bar q}$ structure component of the $\sigma$ meson which is
simpler than the tetraquark structure component and therefore has a larger
transition matrix element. Thereafter a rearrangement of
the structure leads to the tetraquark structure which then decays into
two pions.  In this reaction chain the $q{\bar q}$ structure component
serves as a doorway state for the two-photon excitation of the tetraquark
structure component \cite{schumacher11} . Eq. (\ref{doorway2}) describes
Compton scattering 
via the $t$-channel. In this case only the $q{\bar q}$ structure plays a
r\^ole.
For kinematical reasons the $\sigma$ meson described in Eq. (\ref{doorway2}) 
shows up as a narrow resonance having a definite mass of
$m_\sigma = 666$ MeV, whereas the $\sigma$ meson described in
Eq. (\ref{doorway1}) corresponds to a pole on the second Riemann sheet.

It is apparent that Table \ref{tab:tetraquark} does not contain a scalar meson
having a $s{\bar s}$ structure. This leads to the expectation that one of the
scalar mesons located above 1 GeV should have this structure.
This expectation has been confirmed by
Hatsuda and Kunihiro \cite{hatsuda94} who applied RPA techniques to
the mass relation of the NJL model. In this way it has been predicted
that two $\sigma$ mesons exist, {\it viz.}
\begin{equation}
m_\sigma=668.0\,\, {\rm MeV},\,\,{\rm and}\,\,\, m_{\sigma'}= 1344\,\,{\rm
  MeV} \label{sigmamass}
\end{equation}
where $|\sigma\rangle =|(u{\bar u}+ d{\bar d})/\sqrt{2}\rangle$ and
$|\sigma'\rangle=|s{\bar s}\rangle$. The mass of the $\sigma$ meson
is in close agreement with  the mass $m_\sigma=666$ MeV derived above, showing
that our  
method and the one of Hatsuda and Kunihiro \cite{hatsuda94}
are essentially equivalent.
This gives us confidence that it is appropriate to use $m_{\sigma'}=1344$ MeV
as one  basis for predictions of masses of scalar mesons  containing strange
quarks, in case the effects of the current quarks are included.  
The other basis is the predicted constituent-quark mass in the chiral
limit given in Eq. (\ref{conclusions2}), i.e. for the case that
the effects of the current quarks are not included. In the following we discuss
three models for the mass generation of scalar mesons differing by the
procedure of combining the effects strong-interaction and EW interaction.
These models are extensions of the corresponding models used
in the SU(2) sector.

\subsubsection{First overview on the masses of scalar mesons without
and with including the effects EW interaction}

The masses of scalar mesons may be composed of the masses of the two $\sigma$
mesons making 
contributions in proportion to the fraction of non-strange quarks and
strange quarks, respectively.
The appropriate mass formulae are
\begin{eqnarray}
&&m^{\rm cl}_{\rm scalar}=  (1-f_s) m^{\rm cl}
_\sigma + f_s m^{\rm cl}_{\sigma'},
   \label{scalarmass1}\\
&&m_{\rm scalar}=  (1-f_s) m_\sigma + f_s m_{\sigma'}
\label{scalarmass2}
\end{eqnarray}
where Eq. (\ref{scalarmass1}) refers to the chiral limit where the 
effects of EW interaction are disregarded  and Eq. (\ref{scalarmass2}) 
to the case where these effects are included.
Eq. (\ref{scalarmass1}) is evaluated using $m^{\rm cl}_\sigma=652$ MeV and
$m^{\rm cl}_{\sigma'}=922$ MeV as predicted in section 4.1.
Eq.(\ref{scalarmass2}) is evaluated using the standard value of the mass of the
$\sigma$ meson $m_\sigma=666$ MeV and $m_{\sigma'}=1344$ MeV as predicted by
Hatsuda and Kunihiro \cite{hatsuda94}.
The results of these mass predictions are given in columns 4, (a) and 5, (b) of
Table \ref{scalartable}. By comparing the masses in column (a) with those
of column (b) we see that the largest part of the mass is already present
in the chiral limit, i.e. before EW interaction is taken into account.
The values in column (b) have to be compared with the experimental values
and show a reasonable agreement. 
\begin{table}[h]
\caption{Decuplet of scalar mesons including  $f_0(1370)\equiv 
\sigma'(1344)$. (a) Scalar-meson masses in the chiral limit according to  
Eq. (\ref{scalarmass1}), (b) Scalar-meson masses including the effects
of EW interaction according to Eq. (\ref{scalarmass2}),
 (c) Scalar-meson masses including the effects EW interaction as
 predicted through the masses of pseudo-Goldstone bosons according to 
Eqs. (\ref{ps1}), (\ref{ps2}), (d) Current-quark masses $m^0_s$ of  
the strange quark obtained through adjustments to experimental data according
to Eq. (\ref{ansatz}).  }
\begin{center}
\begin{tabular}{|l|lr|l|l|l|l|l|}
\hline
meson&exp. $m_{\rm scalar}$&ref. &$f_s$&$m^{\rm
  cl}_{\rm scalar}$ (a)&$m_{\rm scalar}$ (b)& $m_{\rm scalar}$ (c)
& $m^0_s$
(d)  \\
\hline
$\sigma(600)$&$600\pm 70$&\cite{schumacher13}&0&$652$&$666$&$666$& $-$\\
$\kappa(800)$& $700 - 900$ &\cite{PDG} &$1/4$&720& 836& 806 & 139 \\
$f_0(980)$,& $980\pm 20$&\cite{PDG}&1/2&787&1005 &929& 191\\
$a_0(980)$&$990\pm 20$& \cite{PDG}&1/2&787&1005&929&191\\
$f_0(1370)$ &$1368\pm 22$&\cite{klempt97}&1&922&1344& $-$&244\\
\hline
\end{tabular}
\end{center}
\label{scalartable}
\end{table}
These experimental values given in Table \ref{scalartable} have been 
obtained as follows. 
The experimental masses of the $\kappa(800)$, $a_0(980)$ and
$f_0(980)$ mesons are taken from the Particle Data Group \cite{PDG}. 
The mass of the  $f_0(1370)$ meson has been determined in three
experiments via the $\bar{p}p\to 5\pi$ reaction \cite{klempt97}, 
leading to the weighted
average listed in line 6 of column 2 in Table \ref{scalartable}.
It is straightforward to identify the predicted scalar meson $\sigma'(1344)$
with the observed 
scalar meson $f_0(1370)$. One argument for this identification
is the agreement of the values obtained
for the masses. An other
argument is that   $f_0(980)$ is excluded because of its tetraquark structure.
A third argument is based on calculations of Fariborz et al. \cite{fariborz15}
leading to arguments in favor of a $s\bar{s}$ structure. One
consequence of these findings is  that the constituent mass of the
$s$ quark including the effects of EW interaction  is rather large, {\it
  viz.} 
\begin{equation}
m_s=\frac12\times 1344 \,\,{\rm MeV}=672 \,\,{\rm MeV}.
\label{smass1}
\end{equation}

\subsubsection{Effects of EW mass generation calculated
 via the masses of pseudo-Goldstone bosons}

In the preceeding subsection the masses of scalar mesons are constructed
for the two cases where the effects of EW interaction
are not included (column (a)) and included (column (b)). 
The considerations presented have the advantage that the effect
of EW interaction on the mass of the scalar meson are clearly
demonstrated and quantitative 
values of the masses for the two cases are predicted.

 A further independent method to take into account the effects of EW interaction
 may be
obtained from the supposition
that
the masses of  the scalar mesons $\kappa(800)$ and 
$f_0(980),\,a_0(980)$ may  be calculated using mass formulae 
analogous to Eq. (\ref{sigma-mass}) which was written down for the $\sigma$
meson. 
These formulae are  
\begin{eqnarray}
&&m_\kappa=\left(\left(m^{\rm cl}_{\rm scalar}(\kappa)\right)^2+\frac12\left(
    m^2_\pi
+m^2_K\right)\right)^{1/2},\label{ps1}\\
&&m_{f_0,a_0}=\left(\left(m^{\rm cl}_{\rm scalar}(f_0,a_0)\right)^2+
(m_K)^2  \right)^{1/2}.\label{ps2}
\end{eqnarray}
In Eqs. (\ref{ps1}) and (\ref{ps2}) 
use is made of
the masses of pseudoscalar mesons $\pi$ and $K$
 in order to take into account the effects of the EW interaction.
These mesons are pseudo-Goldstone bosons and, therefore, have the property
that their masses tend to zero in the chiral limit, as requested in Eqs.
(\ref{ps1}) and (\ref{ps2}).  The results of this calculation are listed in
column 6 (c) of Table \ref{scalartable}. The agreement of the two methods of
calculation given in column  (b) and  (c) of Table \ref{scalartable} is good
for  the $\kappa(800)$ meson but shows the tendency of a deviation in case
of the $f_0(980)$ and $a_0(980)$ mesons. This deviation can be removed
by replacing the K-meson mass by larger value of 590 MeV, showing that 
Eq. (\ref{ps2}) is a reasonable but not perfect approximation.

\subsubsection{Effects of EW interaction taken in account by adding  
current-quark masses to the $\sigma'$ meson mass in the chiral limit}

A further option for the calculation of the EW-interaction part of scalar meson 
masses may be obtained
from the ansatz
\begin{equation}
m_{\sigma'}=m^{\rm cl}_{\sigma'} + 2\,m^0_s
\label{ansatz}
\end{equation}
where  $m^{\rm cl}_{\sigma'}=922$ MeV, as shown in line 6 of column (a)
in  Table
\ref{scalartable}.
This ansatz is an extension of the relation $m_\sigma=m^{cl}_\sigma+
(m^0_u+m^0_d)$ derived in Eq. (\ref{sigmass}).
Here the current-quark mass, $m^0_s$, is an adjustable parameter
determined by adjusting the predicted mass according to Eq. (\ref{scalarmass2})
to the experimental scalar mass. 

The result of this adjustment procedure is shown in column (d) of Table
\ref{scalartable}.
For the meson $\kappa(800)$ the result
$m^0_s=136$ MeV is in good agreement with the the result $m^0_s=147$ MeV
obtained from the mass of $K$ meson (see subsection 4.2). At larger  masses
of the scalar mesons there is a drastic increase of the adjusted
current-quark mass $m^0_s$,  but the results obtained remain in the range
$m^0_s=133$ to $232$ MeV of values found in previous investigations (see
subsection 4.2).

From this
finding the following conclusions may be drawn. First of all there should be
a definite value for the  current-quark  mass of the strange quark, {\it viz.} 
$m^0_s$(EW),       which is the result
of a genuine EW mass production process related to the Higgs boson. 
The different model dependent values $m^0_s$  obtained by adjusting to
experimental data contain additional contributions 
from the $\langle q\bar{q} \rangle$ vacuum condensate which is not taken care
of by the scalar mass calculated for the chiral limit.
These additional contributions may be quite
sizable as shown by the increase of the values in column (d) of Table 
\ref{scalartable} with increasing mass of the scalar meson. An explanation
of this additional contribution would be an interesting topic for further 
investigations.

\subsection{The masses of octet baryons}

Scalar mesons and baryons differ by the fact that scalar mesons are loosely
bound so that the mesons mass can be identified with the sum of
constituent-quark 
masses, whereas  in the case of baryon masses  binding energies
have 
to be taken into account. In Table \ref{tabbaryons} a test of this concept
is carried out. In column (a) the average experimental masses of the three
groups of octet baryons are shown. In columns (b) and (c) the corresponding 
predicted masses in the chiral limit and including the effects of EW interaction
are listed.
\begin{table}[h]
\caption{
(a): average experimental masses of the three
groups of baryons. (b): sum of the masses of
the three 
constituent quarks  in the chiral limit. (c): sum of the masses of the three
constituent quarks including the effects of EW interaction. (d)
    binding energy (B) per number
  of 
quarks (A=3) (in units of MeV).}
\begin{center}
\begin{tabular}{|l|l|l|l|l|}
\hline
baryon & $m_{\rm exp.}$ (a)& $m^{\rm cl}_{\rm theor.}$(b)&$m_{\rm theor.}$
(c)& B/A (d)\\
\hline
$p$, $n$ & 939 &978 & 999& 20\\
$\Lambda$, $\Sigma^{\pm,0}$& 1234 &1113&  1338 & 55\\
$\Xi^{0,-} $& 1318 &1248&  1677 & 120\\
\hline
\end{tabular}
\label{tabbaryons}
\end{center}
\end{table}
Apparently these predicted masses of column (c) 
are larger than the experimental ones,
confirming the expectation that binding energies are involved in the 
formation of the total baryon mass. The binding energies  $B$ per number $A=3$
of 
constituent quarks  are listed in column (d). These binding energies 
increase with the fraction of strange quarks located in the baryon. This
interesting finding needs an explanation.

\subsection{The magnetic moments of Hyperons}

The magnetic moments of hyperons have been predicted making use of different
assumptions about the masses of the constituent quarks (see e.g. 
\cite{scadron06}). Within the quark model the masses of the constituent quarks
are the only parameters in the prediction of magnetic moments. 
Therefore, within this model information on the
constituent-quark masses is obtained by comparing predicted magnetic moments
with experimental values.

In the present investigation we use $m_u=331$ MeV,
$m_d=335$ MeV and $m_s=672$ MeV. 
\begin{table}[h]
\caption{Constituent mass of strange quark calculated from magnetic moments}
\begin{center}
\begin{tabular}{|l|c|l|r|}
\hline
hyperon& $(9/m_p)\times\mu^{\rm theor.}$ &$\mu_{\rm exp. }$&$\mu^{\rm theor.}$\\
\hline
$\Lambda$&$ -3/m_s$& $-0.613\pm0.004$&$-0.465$\\
$\Sigma^+$&$8/m_u+1/m_s$& $2.458\pm 0.010$&$2.675$\\
$\Sigma^-$& $-4/m_d+1/m_s$ & $ -1.160\pm 0.025$&$-1.090$\\
$\Xi^0$&$-4/m_s-2/m_u$&$-1.250\pm 0.014$&$-1.250$\\
$\Xi^-$& $-4/m_s+1/m_d$&$-0.6507\pm 0.0025$&$-0.309$\\
$\Omega^-$&$-9/m_s$&$-2.02\pm 0.05$&$-1.396$\\
\hline
\end{tabular}
\end{center}
\label{tab:magmoments}
\end{table}
With these constituent-quark masses the predicted magnetic moments $\mu^{\rm
theor.}$ given  in column 4 of Table \ref{tab:magmoments} are obtained.
The value predicted for the $\Xi^0$ hyperon is in perfect agreement with
experimental value. In other cases the agreement is less pronounced. 
However, by using a different choice  for  the constituent mass of the strange
quark no improvement of the general agreement is obtained.
The conclusion we have to draw is that our set of constituent-quark masses
is in line with the experimental magnetic moments, though there is no  strong
support 
for  the specific value adopted  for the $s$ quark. 
The information obtained from the masses of
scalar mesons on the constituent mass of the $s$ quark
  appears to be more reliable than the one obtained from magnetic moments.

\section{Summary and Conclusions}

\begin{figure}[h]
\begin{center}
\includegraphics[width=0.44\linewidth]{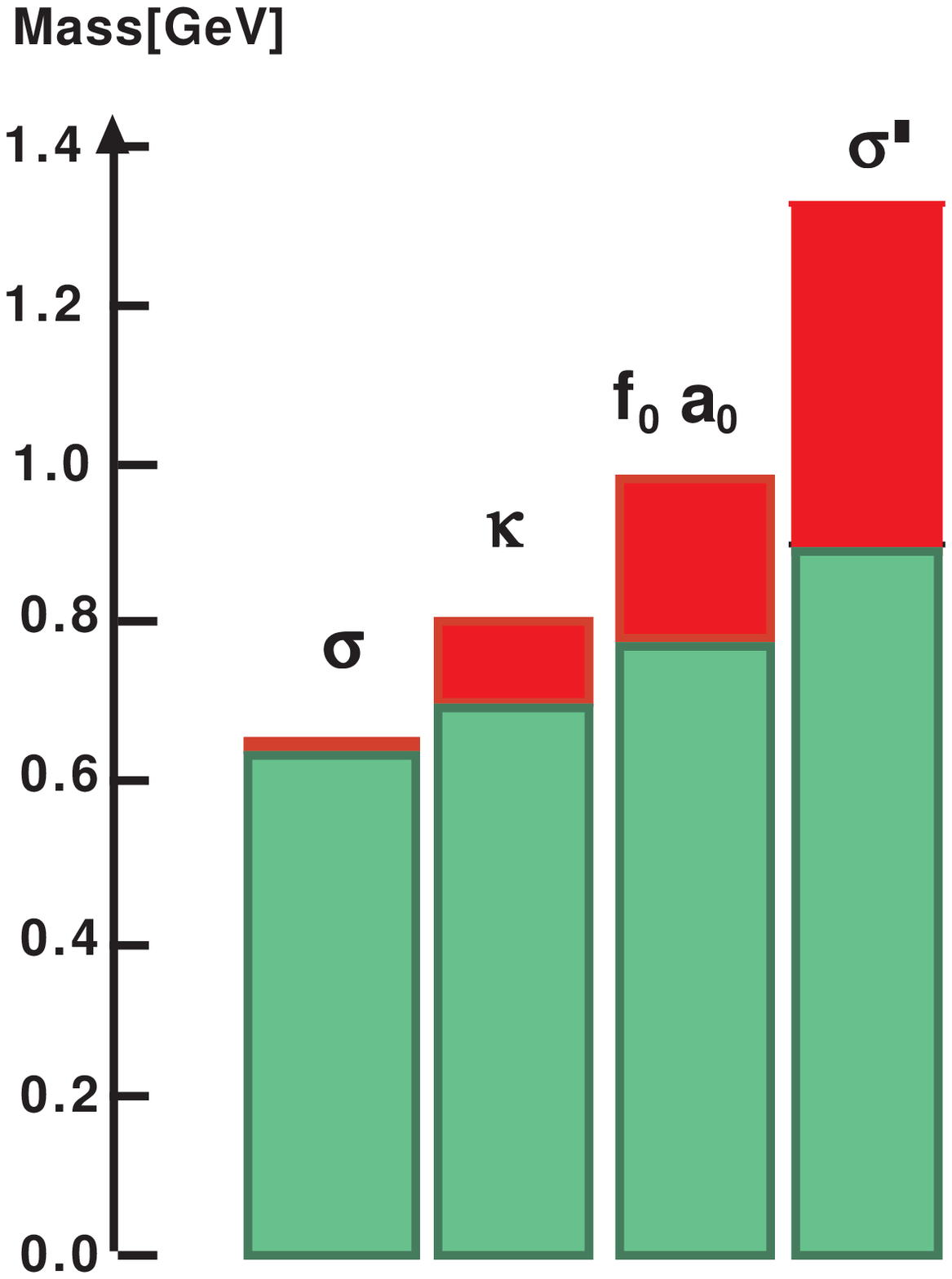}
\includegraphics[width=0.55\linewidth]{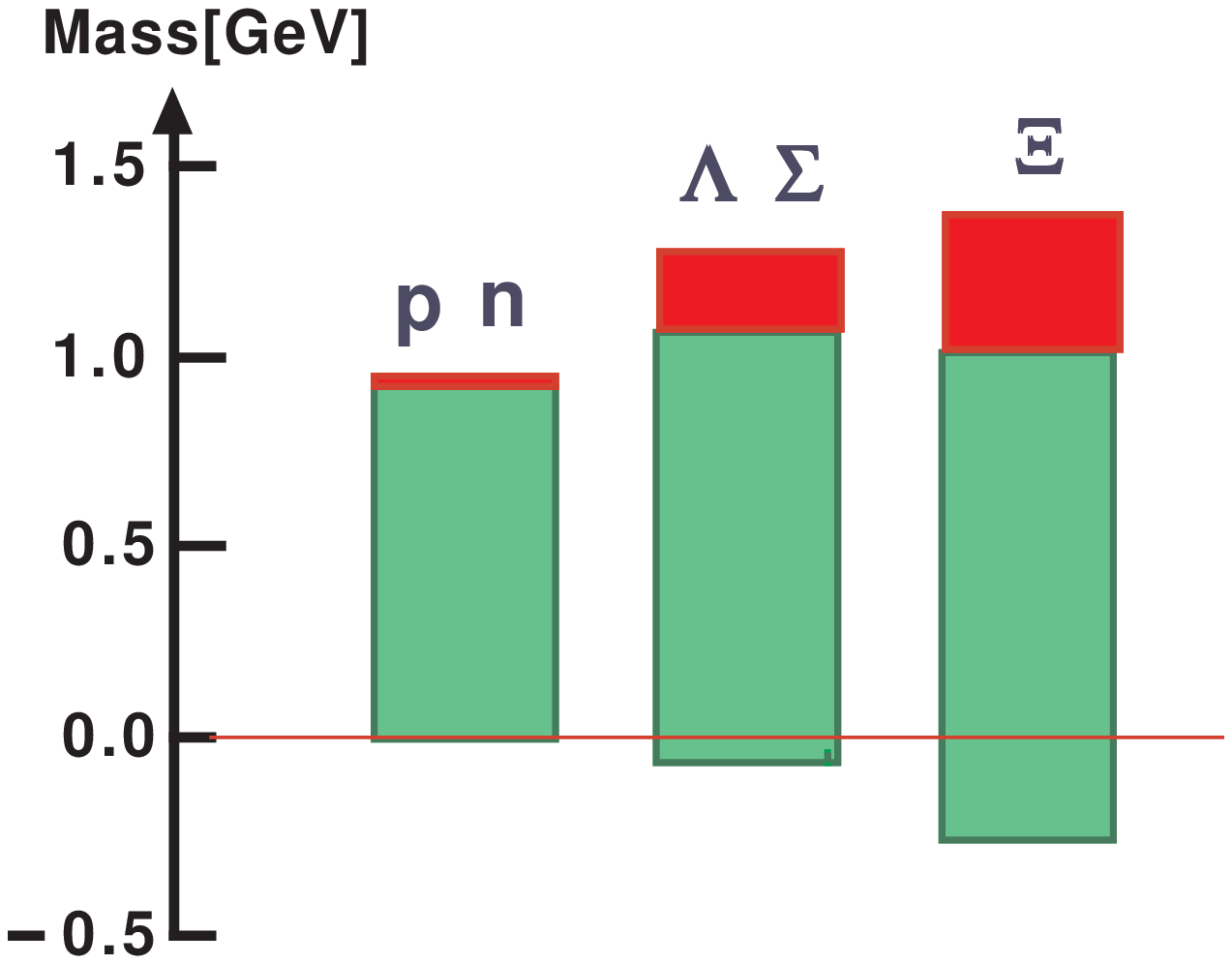}
\end{center}
\caption{Masses generated via strong-interaction (green or light-grey boxes) 
and EW interaction (red or dark-grey boxes).  Left panel: Scalar mesons.
Right panel:
Octet baryons. The negative parts of the green boxes in the right panel 
correspond
the binding energies observed in the octet baryons. Without the effects of the
Higgs boson only the masses represented by the green boxes would be present.
The red boxes correspond to the effects of the Higgs boson.}
\label{smassne}
\end{figure}
In the forgoing it has been shown that the prediction of scalar-meson
and baryon masses can be extended from the SU(2) sector to the SU(3)
sector by making use of a second $\sigma$ meson, {\it viz.}
 $\sigma'(1344)$, having
the $|s\bar{s}\rangle$ structure. This second $\sigma$ meson implies
that the constituent-quark mass of the $s$ quark when including the effects  
of the Higgs boson is $m_s=672$ MeV. This large constituent-quark mass leads to 
reasonable predictions of  the  masses of scalar mesons below 1 GeV 
and of the masses  
of octet baryons. Furthermore, also the magnetic moments of octet baryons are
predicted 
leading to results being in  line with the experimental values.

The masses of constituent quarks are composed of the masses $M_q$ predicted
for the chiral limit and the mass of the 
respective current quark $m^0_q$
provided by the Higgs boson (EW interaction) alone. 
For scalar mesons the sum 
of $M_q$ and
$m^0_q$ leads to a zero-order approximation for the constituent-quark
mass $m_q$, but there  are dynamical effects described by the NJL model which
modify the simple relation $m_q= M_q + m^0_q$, except for the non-strange
sector where this relation is  a  good approximation. Similar results are
obtained for the octet baryons. A difference between the scalar mesons and the
octet baryons is that that for scalar mesons binding energies do not play a
r\^ole whereas they are of importance in case of octet baryons. 

In Figure \ref{smassne} a graphical a representation of  mass generation 
is given. The green boxes correspond to masses in a world without the Higgs
boson, whereas the red boxes represent the effects of the  Higgs boson
on the  mass generation process.

\end{document}